\documentclass[preprint,aps,floatfix,onecolumn]{revtex4}
\usepackage{epsfig,amsmath,bbm,graphicx,color}

\begin{document}

\title{New method for deciphering free energy landscape of three-state proteins} 

\author{Mai Suan Li$^1$, A. M. Gabovich$^2$, and A. I. Voitenko$^2$}

\affiliation{$^1$Institute of Physics, Polish Academy of Sciences,
Al. Lotnikow 32/46, 02-668 Warsaw, Poland\\
$^2$Institute of Physics, Nauka Avenue 46, Kiev 03680, Ukraine}

\begin{abstract}
We have developed a new simulation method to estimate the distance
between the native
state 
 and the first transition state, and the distance  between
the intermediate state and the second transition state  of a
protein which mechanically unfolds via intermediates. Assuming that
the end-to-end extension $\Delta R$ is a good reaction coordinate
to describe
the free energy landscape of proteins subjected to an external force,
we define the midpoint extension $\Delta R^*$ between two transition states
from either constant-force or constant loading rate pulling
simulations. In the former case,
 $\Delta R^*$ is defined as a middle
point between two plateaus in the time-dependent curve of $\Delta R$,
while, in the latter one, it is a middle point between two peaks in
the force-extension curve.
Having determined $\Delta R^*$, one can compute times needed to cross two
transition state barriers starting from the native state.
With the help of the Bell and microscopic kinetic theory,
force dependencies of these unfolding times can be used to locate the
intermediate state and to extract unfolding barriers.
We have applied our method to the titin domain I27 and
the fourth domain of {\em Dictyostelium discoideum} filamin
(DDFLN4), and obtained reasonable agreement with experiments,
using the C$_{\alpha}$-Go model.
\end{abstract}


\maketitle

\vfill

\section{Introduction}

Despite numerous advances achieved within recent years
\cite{Onuchic_COSB04},
deciphering the free energy landscape (FEL)
of biomolecules remains a
challenge to molecular biology. The most detailed information on the FEL
may be gained
from all-atom simulations, but this approach,
due to its high computational expenses, is restricted to rather
short peptides and proteins \cite{Gnanakaran_COSB03,Phuong_Proteins05}.
In this situation, the
AFM \cite{Rief_Science07} or optical
tweezers \cite{Kellermayer_Science97} have been proved a very useful tool 
for  probing the FEL of proteins.
In most experiments
\cite{Brockwell_BJ02,Carrion-Vazquez_PBMB00}, assuming
that  mechanical unfolding is well described by a two-state scenario,
one can extract the distance $x_u$ between the native state (NS) and the transition
state (TS), as well as the unfolding barrier $\Delta G^{\ddagger}$.
Latest theoretical studies \cite{West_BJ06,MSLi_BJ07a,Kouza_JCP08} showed that simple
coarse-grained models can provide reasonable estimates for those quantities.

\begin{figure}
\epsfxsize=6.3in
\vspace{0.2in}
\centerline{\epsffile{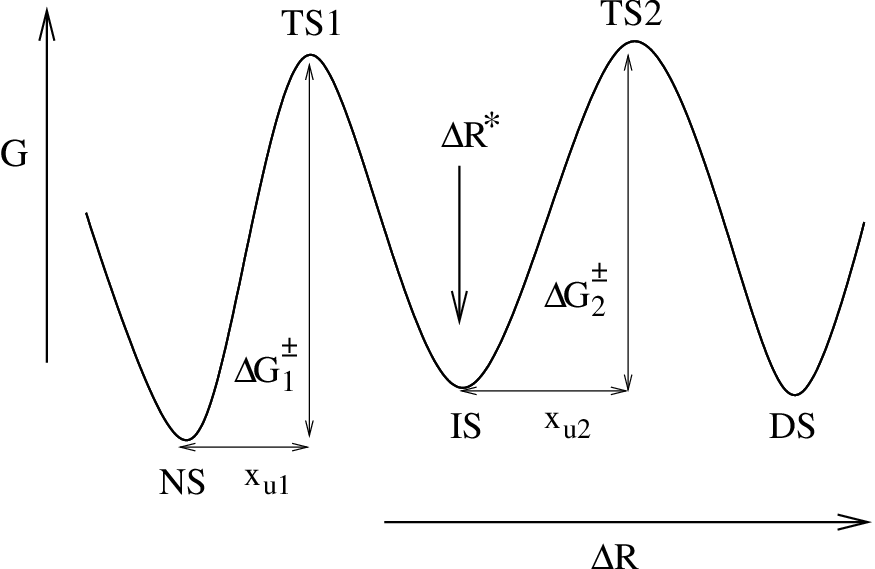}}
\caption{}
\label{free_3state_concept_fig}
\end{figure}

In recent experimental works \cite{Williams_Nature03,Schwaiger_EMBO05},
 the FEL of three-state proteins
has been
studied. A schematic plot of the FEL for this class of proteins is
shown in Fig. \ref{free_3state_concept_fig}. Single molecule force
measurements allow  
the distance $x_{u1}$ between the native and the first transition
state (TS1) and the distance $x_{u2}$ between the intermediate state (IS)
and the second transition state (TS2) to be evaluated. The unfolding barriers
$\Delta G^{\ddagger}_1 = G_{TS1} - G_{NS}$ and
 $\Delta G^{\ddagger}_2  = G_{TS2} - G_{IS}$
can also be determined.
As far as we know, there were no theoretical
attempts to calculate these important parameters. Therefore, the
goal of this work is to develop a new method  for
computing them from simulations.
Using the
Go model \cite{Clementi_JMB00} and our new
method,
we calculated the free energy landscape parameters
for the three-state titin domain I27
and
the domain DDFLN4.
Our results
are in reasonable agreement with the experiments \cite{Williams_Nature03,Schwaiger_EMBO05}.

\section{Method}

The new method is based on the fact that
the end-to-end extension, $\Delta R$, is a good reaction coordinate
for describing mechanical unfolding.
It should be noted that the direction of the force and
therefore $\Delta R$ is changing upon partial unfolding of a
protein.
The time $\tau_{u1}$ for a molecule
to go from the NS to the IS is defined as a time needed
to achieve the intermediate point $\Delta R^*$ between the TS1 and TS2
(Fig. \ref{free_3state_concept_fig}).
The time
to cross the TS2 barrier starting from the IS, $\tau_{u2}$, is a time
to reach the rod state starting from $\Delta R^*$.
We propose to determine $\Delta R^*$ from
either constant-force or constant loading rate pulling
simulations.
Using the force dependencies of $\tau_{u1}$ and $\tau_{u2}$, and the Bell
formula \cite{Bell_Sci78}, we can extract $x_{u1}$ and $x_{u2}$.
The unfolding barriers $\Delta G^{\ddagger}_1$ and $\Delta G^{\ddagger}_2$
can be estimated in the framework of the microscopic kinetic theory
\cite{Dudko_PRL06}.

To illustrate our method, we used
the
Go model \cite{Clementi_JMB00}, which
is an appropriate choice
not only because
the construction of FEL for long enough biomolecules by all-atom models
is beyond present computational facilities, but also
because unfolding
of biomolecules is mainly governed by the
native conformation topology
\cite{West_BJ06}. Moreover,
it has been recently shown \cite{MSLi_BJ07a,Kouza_JCP08}
that 
the Go model
provides
reasonable estimates for $x_u$ and $\Delta G^{\ddagger}$
in the two-state approximation
 and  is
expected to be applicable to three-state
molecules.
Details of the Go model are given in our previous works 
\cite{MSLi_BJ07a,Kouza_JCP08}. We
use the same set of parameters as for Go modeling of ubiquitin
\cite{MSLi_BJ07}.
 The main computations were carried out at $T = 285$ K
$= 0.53 \epsilon _H/k_B$, where 
$k_B$ is the Boltzmann constant, and $\epsilon _H = 0.98 \, {\rm kcal/mol}$
is the  hydrogen bond energy.
The friction $\gamma _0$ was chosen to be the same as for water,
i.e. $\gamma_0 = 50 \frac{m_a}{\tau_L}$
\cite{Veitshans_FoldDesign97}, where
 $\tau_L = (m_aa^2/\epsilon_H)^{1/2} \approx 3$ ps. Here
the characteristic bond length between successive beads
$a \approx 4 \AA \,$ and  the typical mass of amino acid residues
$m_a \approx 3\times 10^{-22}$ g \cite{Veitshans_FoldDesign97}.
For the water viscosity, one can neglect the inertia term and use the Euler method
with the time step $\Delta t = 0.1 \tau_L$
to solve the corresponding Langevin equation.

Since native titin is a highly heterogeneous polymer,
in experiments, one used a poly-protein composed of 
identical tandem repeats of the
Ig module (I27$_{12}$ \cite{Carrion-Vasquez_PNAS99,Marszalek_Nature99,Williams_Nature03}) 
to study elastic properties of a single domain at high solution.
In the DDFLN4 case, a single DDFLN4 domain is sandwiched between Ig domains 
I27-30 and domains I31-34 from titin \cite{Schwaiger_NSB04}.
Unfolding of DDFLN4 can be easily studied, as its mechanical stability
is much lower than that of Ig-domains (DDFLN4 would have lower
peaks in the force-extension curve).
In AFM experiments, one end of a poly-protein is anchored to a surface
and the another end to a tip of
a very soft cantilever.
The molecule is stretched by increasing the distance
between the surface and the cantilever as the external force acts on
one of termini via the tip. Because a poly-protein mechanically unfolds
domain by domain, one can consider that one end of a single domain
is anchored and the force is applied to the another one.
Therefore, as in all-atom steered molecular dynamics
simulations \cite{Lu_BJ98}, when the force is ramped linearly with time,
we fix the N-terminal and pull the
C-terminal by applying a force $f = K_r(vt -r)$. 
Here $r$ is the displacement of the pulled atom 
from its original position
\cite{Lu_BJ98}, and the spring constant  $K_r$ is set to be the same as in
the harmonic term of the potential energy for the studied system
\cite{Clementi_JMB00,MSLi_BJ07}.
The pulling direction was chosen along the vector directed
 from the fixed atom to the
pulled one. The pulling speed was set equal to
$v = 3.6\times 10^7$ nm/s, which is
about 3 - 4 orders of magnitude faster than those used in experiments.
In the constant force simulations, we add the term
$-\vec{f}\vec{R}$ to the total energy of the system,
where $R$ is the end-to-end distance, and $f$ the force applied
to the both termini.

We define the unfolding time $\tau_{u1}$
as an average of the first passage times needed to reach the extension
$\Delta R^*$.
Different trajectories start from the same native
conformation, but with different random number seeds.
The unfolding time $\tau_{u2}$ is an average of the
first passage times needed for the molecule
to achieve the  rod state starting from the extension
$\Delta R^*$.
In order
to get a reasonable estimate for $\tau_{u1}$ and $\tau_{u2}$,
we have generated 30 - 50 trajectories
for each value of $f$.

\section{Results}

The crucial point in our method is how to determine $\Delta R^*$. 
We will show that both simulation ways specified above are valid
for this purpose.\\

{\bf Determination of $\Delta R^*$ for I27}\\

\begin{figure}
\epsfxsize=6.3in
\vspace{0.2in}
\centerline{\epsffile{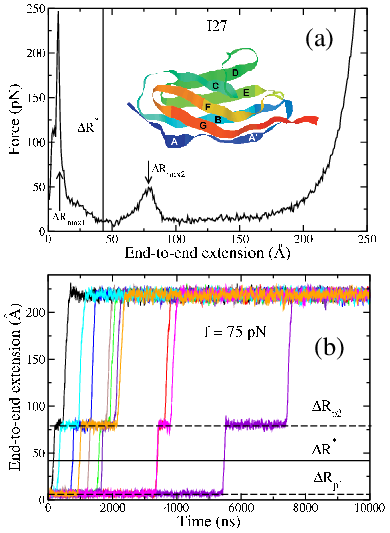}}
\caption{}
\label{f_ext_v5_snap_titin_fig}
\end{figure}

Fig. \ref{f_ext_v5_snap_titin_fig}a presents the force-extension curve obtained
at the speed $v = 3.6\times 10^7$ nm/s.
In accordance with the experiments \cite{Marszalek_Nature99} and
all-atom simulations \cite{Lu_BJ98},
we observe two peaks, which ascertain that unfolding proceeds via
intermediates (if the protein
unfolded without intermediates, a single peak would be observed).
The first peak, located at $\Delta R_{max1} \approx 8$ \AA ,
occurs due to a detachment of strand A (Fig. \ref{f_ext_v5_snap_titin_fig}a)
from the protein core.
One can show that the appearance of the second peak at $\Delta R_{max2} \approx 78$ \AA $\;$
is related to full unfolding of strands A', F, and G.
Assuming that $\Delta R^*$ is a middle point between two local peaks,
we have $\Delta R^* = (\Delta R_{max1} + \Delta R_{max2})/2 \approx 43$ \AA.

We now intend to show that $\Delta R^*$ can also be determined from
constant-force simulations.
As is evident from Fig. \ref{f_ext_v5_snap_titin_fig}(b),
two
plateaus occur at $\Delta R_{p1}
 \approx 6$  \AA $\;$
and $\Delta R_{p2} \approx 78$ \AA $\;$  in the time 
dependence of $\Delta R(t)$.
Within error bars, locations of plateaus coincide with those for
peaks shown in Fig. \ref{f_ext_v5_snap_titin_fig}(a).
Again, the occurrence
of two plateaus indicates that this domain mechanically unfolds in
a three-state manner.
Since the plateaus are related to crossing the unfolding barriers
$\Delta G^{\ddagger}_1$ and $\Delta G^{\ddagger}_2$,
the middle point $\Delta R^*$ between the
TS1 and TS2 can be identified
as a middle point between two plateaus, i.e.
$\Delta R^* \approx (\Delta R_{p1} + \Delta R_{p2})/2$. For I27,
$\Delta R^* \approx 42$ \AA, which is close to the result,
obtained by constant-velocity pulling simulations. \\

{\bf Determination of $\Delta R^*$ for DDFLN4}\\

\begin{figure}
\epsfxsize=6.3in
\vspace{0.2in}
\centerline{\epsffile{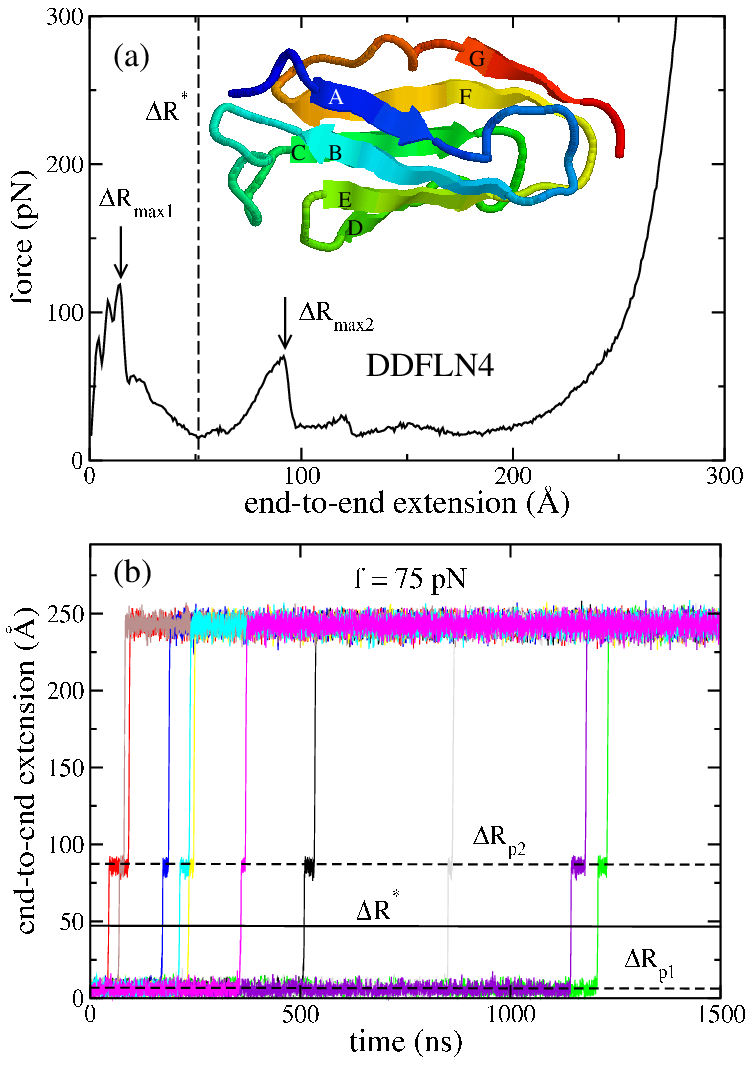}}
\caption{}
\label{ddfln4_force_ext_ns_fig}
\end{figure}

Fig. \ref{ddfln4_force_ext_ns_fig}b shows the force-extension profile
for the same pulling speed as in the I27 case.
Two peaks 
appear
at $\Delta R_{max1} \approx 14$ and $\Delta R_{max2} \approx 92$ \AA.
Using these values, we obtain  
$\Delta R^* = (\Delta R_{max1} + \Delta R_{max2})/2 = 53$ \AA.
The existence of intermediates is also evident from the constant force
simulations which give two plateaus 
at $\Delta R_{p1}
 \approx 7$  \AA $\;$
and $\Delta R_{p2} \approx 85$ \AA $\;$ (Fig. \ref{ddfln4_force_ext_ns_fig}b). 
Therefore,
$\Delta R^* = (\Delta R_{p1} + \Delta R_{p2})/2 \approx 46 \AA \;$ which
is not far from the value followed from the force-extension curve.
Since
the peaks are not sharp and
the plateau position fluctuates,
one can consider that two simulation modes gave the same result.
We will use the averaged value $\Delta R^*=50 \AA\;$ for computing
$\tau_{u1}$ and $\tau_{u2}$ for DDFLN4.

\begin{figure}
\epsfxsize=6.3in
\vspace{0.2in}
\centerline{\epsffile{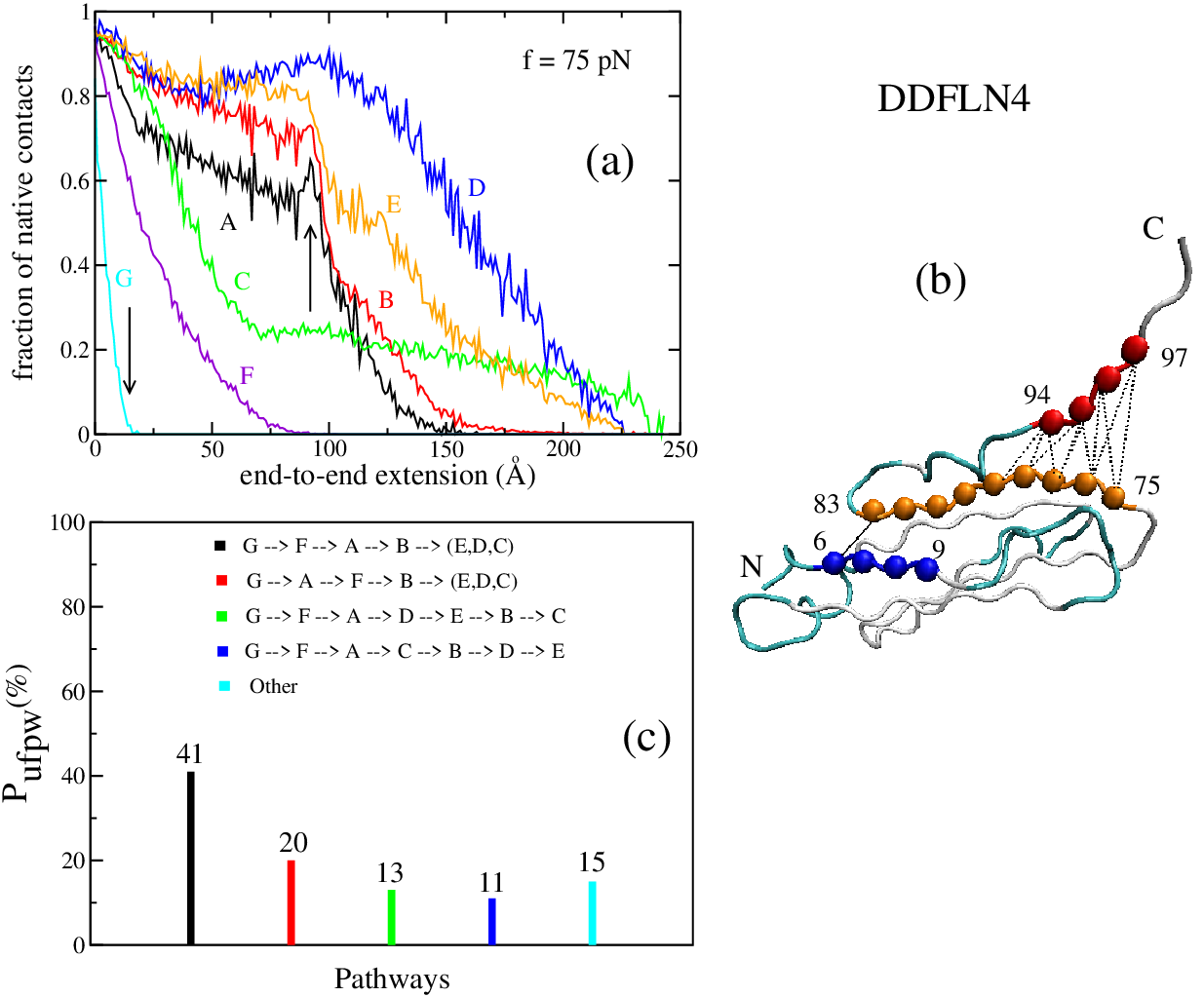}}
\caption{}
\label{ddfln4_cont_ext_pathway_snap_fig}
\end{figure}

In accordance with the experiments \cite{Schwaiger_NSB04,Schwaiger_EMBO05},
the Go model captures the overall behavior of
DDFLN4 that it mechanically unfolds via intermediates.
However, the location and structure of Go intermediates are different
from the experimental
ones. In the experimental force-extension
curve \cite{Schwaiger_NSB04},
 two peaks occurs at $\Delta R_{p1} \approx 150 \AA \;$ and
$\Delta R_{p2} \approx 310 \AA \;$ which are very different from our
results (Fig. \ref{ddfln4_force_ext_ns_fig}a). 
Using loop mutations \cite{Schwaiger_NSB04}, it was suggested
that during the first unfolding event (first peak) strands A and  B
detach from the domain and unfold.
Therefore, strands C - G form a stable intermediate structure, which then
unfolds in the second unfolding event (second peak).
In order to sort out intermediates in the Go model,
 we plot fractions
of native contacts formed by each strand with the rest of the protein
as a function of $\Delta R$ (Fig. \ref{ddfln4_cont_ext_pathway_snap_fig}a).
Here, we present the results obtained for the case when 
the N-terminal is kept fixed and
the force is applied to the C-terminal with the same loading rate
as in Fig. \ref{ddfln4_force_ext_ns_fig}a. 
%
%
At the position of the first peak in the force-extension curve
($\Delta R = 14 \AA\;$), strand G fully
unfolded, while strands A and B are still structured
(Fig. \ref{ddfln4_cont_ext_pathway_snap_fig}a). Thus, contrary to the
experiments \cite{Schwaiger_NSB04,Schwaiger_EMBO05}, Go intermediate
 conformations
consist of six strands A-F. A typical snapshot of intermediates is shown
in Fig. \ref{ddfln4_cont_ext_pathway_snap_fig}b, where all contacts between
G and F are broken, but a single contact between A and F remains intact.
At the second peak position 
($\Delta R = 92 \AA\;$) denoted by an arrow in Fig. \ref{ddfln4_cont_ext_pathway_snap_fig}a, together with G, strand F becomes unstructured
 and most of contacts of strand
C are lost. The number of contacts of strands A, B, D and E
drops drastically as the protein unfolds quickly after this second unfolding
event. As in the experiments \cite{Schwaiger_NSB04,Schwaiger_EMBO05},
at $\Delta R \approx 150 \AA\;$, strands A and B are detached from the core,
but in our Go model, strands F and G  have already  unfolded.

From Fig. \ref{ddfln4_cont_ext_pathway_snap_fig}a, we obtain
the following unfolding pathway for DDFLN4:
\begin{equation}
\textrm{G} \rightarrow \textrm{F} \rightarrow \textrm{A} \rightarrow 
\textrm{B} \rightarrow \textrm{(E,D,C)}.
\label{unfolding_pathways_eq}
\end{equation}
In addition to this dominant pathway, other routes to the rode state are also
possible (Fig. \ref{ddfln4_cont_ext_pathway_snap_fig}c). Mechanical unfolding
pathways may be different, but they share a common feature that strand
G always unfolds first. This also contradicts the experimental suggestion
\cite{Schwaiger_NSB04} that unfolding initiates from the N-terminal.
It is not entirely clear, why the Go model gives different unfolding
pathways and intermediates compared to the experiments. Presumably,
the discrepancy comes from the simplification of Go modeling, where the
non-native interaction and the effect of environment are omitted. 
All-atom simulations are required to clarify this issue.

{\bf Calculation of free energy landscape parameters}\\

\begin{figure}
\epsfxsize=5.5in
\vspace{0.2in}
\centerline{\epsffile{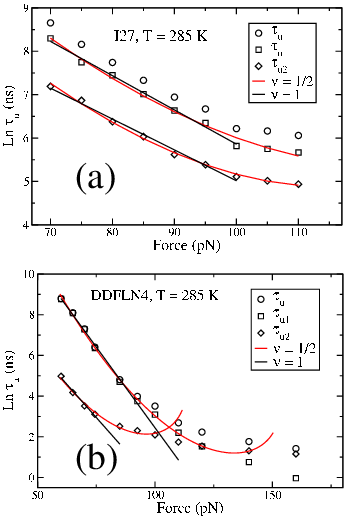}}
\caption{}
\label{x_barrier_I27_ddfln4_fig}
\end{figure}

Once $\Delta R^*$ is found, one can compute the times 
$\tau_{u1}$ and $\tau_{u2}$ as the functions of the external force $f$ and
extract $x_{u1}$ and $x_{u2}$, using the Bell equation \cite{Bell_Sci78}:
\begin{equation}
 \tau_{ui} = \tau_{ui}^0\exp(-fx_{ui}/k_BT), i = 1,2.
\label{Bell_eq}
\end{equation}
We have tried several values of $\Delta R^*$ in the interval
$\Delta R^* = (42 \pm 15)$ \AA, and $\Delta R^* = (50 \pm 15) \AA\;$
for I27 and DDFLN4, respectively. Since the results
remain essentially the same,
we will present those obtained for $\Delta R^* = 42$ \AA,
and $\Delta R^* = 50 \AA\;$ for I27, and DDFLN4, respectively.
Fig. \ref{x_barrier_I27_ddfln4_fig} shows the force dependencies
 not only for
$\tau_{u1}$ and $\tau_{u2}$, but also for the full unfolding time
$\tau_{u} = \tau_{u1} + \tau_{u2}$.
{\em
Strictly speaking, the formula $\tau_{u} = \tau_{u1} + \tau_{u2}$
is valid if the probability of missed unfolding events is negligible.
This happens when the applied force exceeds several pN,
but not in the $f \rightarrow 0$ limit \cite{Schlierf_BJ06}.
Since our computations were carried out at $f$ of tens pN
(Fig. \ref{x_barrier_I27_ddfln4_fig}), the mentioned above
equality is applicable to extract $\tau_{u}$.
}
%

For I27, $\tau_{u1}$ is about 2-3 times
larger than $\tau_{u2}$. It is also evident from
 Fig. \ref{f_ext_v5_snap_titin_fig}(b),
which demonstrates that the second plateau exists during shorter time
 intervals 
than the first one.
A similar situation happens for DDFLN4, but, for high forces,
$\tau_{u2}$ becomes
eventually larger than $\tau_{u1}$ (Fig. \ref{x_barrier_I27_ddfln4_fig}(b)).

In the low-force regime, fitting the data by Bell Eq.
(\ref{Bell_eq}) (straight lines in Fig. \ref{x_barrier_I27_ddfln4_fig}),
we obtain
$x_{u1}$, and $x_{u2}$ quoted in Table.
For both I27 and DDFLN4
our estimate of $x_{u2}$
agrees very well with the experiments, while that for $x_{u1}$ is
a bit higher than the experimental  data.
Given the simplicity of the Go model, the agreement between
the theory and the experiment
should be considered reasonable, but it would be interesting to check if
a more comprehensive
account for non-native contacts and environment could improve
our results.

It is to be noted here that
although the Go model
gives a different location of  intermediates
in the DDFLN4 force-extension curve in comparison with the experiments, it still
provides reasonable estimates for $x_{u1}$ and $x_{u2}$. This is because
the nature of $x_{u1}$ and $x_{u2}$ is different from that
of the end-to-end distance: the former are a measure of the force dependence
of barrier crossing rates, while the later is a real distance.
Nevertheless, one has
to be careful in comparison of Go results with experiments on DDFLN4.

In the Bell approximation, one assumes that the location
of the transition state does not move under the action of an
external force. However,
our simulations for ubiquitin, e.g., showed that it does move toward
the NS \cite{MSLi_BJ07}. 
Recently, assuming that $x_u$ depends on the external force
and using the Kramers theory, 
several groups \cite{Schlierf_BJ06,Dudko_PRL06}
have tried to go beyond the Bell approximation. We follow 
Dudko {\em et al.} who proposed
the following force dependence for the unfolding time \cite{Dudko_PRL06}:
\begin{eqnarray}
\tau _u \; = \; \tau _u^0
\left(1 - \frac{\nu x_u}{\Delta G^{\ddagger}}\right)^{1-1/\nu}
\exp\lbrace -\frac{\Delta G^{\ddagger}}{k_BT}[1-(1-\nu x_uf/\Delta G^{\ddagger})^{1/\nu}]\rbrace.
\label{Dudko_eq}
\end{eqnarray}
Here, $\Delta G^{\ddagger}$ is the unfolding barrier, and $\nu = 1/2$ and 2/3
for the cusp \cite{Hummer_BJ03} and the
linear-cubic free energy surface \cite{Dudko_PNAS03}, respectively.
Note that
$\nu =1$ corresponds to the phenomenological
Bell theory (Eq. (\ref{Bell_eq})).
An important consequence following from
Eq. (\ref{Dudko_eq}), is that one can apply it to estimate not only
$x_u$, but also $G^{\ddagger}$, if $\nu \ne 1 $.
Since the fitting with $\nu = 1/2$ is valid in a wider interval
as compared to the $\nu = 2/3$ case, we
consider the former case only.
The region,
where the $\nu = 1/2$ fit works well, is expectedly wider  than that for
the Bell scenario (Fig. (\ref{x_barrier_I27_ddfln4_fig})). However, for DDFLN4 this fit
can not cover the entire
force interval. 

In the I27 case, from 
the nonlinear fitting (Eq. (\ref{Dudko_eq}) and
 Fig. \ref{x_barrier_I27_ddfln4_fig}(a)),
we obtain $ x_{u1} = 4.7$ \AA, and $x_{u2} = 5.1$
 \AA $\;$, which
are larger than the Bell theory-based estimations (see Table). Using 
raw experimental data \cite{Carrion-Vasquez_PNAS99}
and fitting with $\nu = 1/2$, in two-state approximation,
 Dudko {\em et al.} \cite{Dudko_PRL06} obtained
$x_u = 4 \AA;$, which is close to our result.
For DDFLN4, the
nonlinear fit (Fig. \ref{x_barrier_I27_ddfln4_fig}(b))
gives $ x_{u1} = 13.1$ \AA, and $x_{u2} = 12.8
 \AA\,$ which
are about twice as large as the Bell theory-based estimates (Table).
Such high values of $x_u$ are typical for $\alpha$-proteins
and they may come from the low resistance
of DDFLN4 because the less stable  protein, the larger is $x_u$
\cite{MSLi_BJ07a}. Recently, Dietz and Rief \cite{Dietz_PRL08} have shown
that the product $f_ux_u \approx 50$ pN nm is probably
universal for all proteins.
 Using the experimental
result $f_u \approx 45$ pN \cite{Schwaiger_NSB04,Schlierf_BJ06} and $x_u \approx 13 \AA\;$, we obtain
$f_ux_u \approx 59$ pN nm which is not far from this universal value. 
From this point of view, big values of $ x_{u1}$ and $x_{u2}$ are still
acceptable,  
but additional experimental data are required to settle this problem.

Theoretical values for $G^{\ddagger}_1$, and $G^{\ddagger}_2$,
followed from Fig. \ref{x_barrier_I27_ddfln4_fig},
are listed in Table.  
To estimate the unfolding barriers from the available
experimental data for I27 \cite{Williams_Nature03} and
DDFLN4 \cite{Schwaiger_EMBO05,Schlierf_BJ06},
we used the
following formula:
\begin{equation}
\Delta G^{\ddagger} = -k_BT\ln(\tau _A/\tau _u^0),
\label{UnfBarrier_eq}
\end{equation}
where $\tau _u^0$ denotes the unfolding time in the absence of force, and
$\tau _A$ is a typical unfolding prefactor. Since $\tau _A$ is
not known for unfolding, we used the value typical of folding,
$\tau _A = 0.1 \mu$s
\cite{MSLi_Polymer04,Schuler_Nature02,Schlierf_JMB05}.

For I27, we used $\tau_{u2}^0 = (3)^{-1}\times 10^4 $s and total unfolding time
$\tau_{u}^0 = (2.9)^{-1}\times 10^4 $s \cite{Williams_Nature03}.
$\tau_{u1}^0$ is extracted as $\tau_{u1}^0 = \tau_{u}^0 - \tau_{u2}^0$.
Using these unfolding times and Eq. (\ref{UnfBarrier_eq}),
$G^{\ddagger}_1$, and $G^{\ddagger}_2$ were calculated 
(Table). The best agreement between theory and experiment is obtained
for $G^{\ddagger}_2$. Interestingly, using the similar
fitting procedure and raw experimental data \cite{Carrion-Vasquez_PNAS99},
in two-state approximation,
Dudko {\em et al.} \cite{Dudko_PRL06} obtained $G^{\ddagger} = 20 k_BT$,
which is close to our result.

 In the DDFLN4 case, we used $\tau_{u1}^0 = (0.28)^{-1} $s and
$\tau_{u2}^0 = (0.33)^{-1} $s of \cite{Schwaiger_EMBO05} to estimate
$G^{\ddagger}_1$ and $G^{\ddagger}_2$.
Our result for $\Delta G^{\ddagger}_2$ agrees well with the 
experimental data,
but the theoretical value for $\Delta G^{\ddagger}_1$
turned out
higher than the experimental one. This disagreement may be due to
the limitation of the Go model.
An another possible reason is that the experimental estimations
were obtained using the same prefactor $\tau _A = 0.1 \mu$s for all cases
and this might be invalid.


To conclude,
we have proposed a new simulation
approach to delineate the FEL of multi-state proteins. 
Our method is simple to use, and it does not require any extra CPU cost, because
the unfolding times $\tau_{u1}$, and $\tau_{u2}$ are computed
in a single run for every trajectory.
Using this method and the simple Go model,
we obtained $x_{u1}$, and $x_{u2}$,
which are 
in reasonable agreement with the experimental data for I27
\cite{Williams_Nature03} and DDFLN4 \cite{Schwaiger_EMBO05,Schlierf_BJ06}.
 There is a
discrepancy between theoretical and experimental estimations for some
unfolding barriers. Therefore,
it would be useful to go beyond the Go model to see if one could obtain
better agreement.
Our method is universal and 
may be applied to other multi-state biomolecules. The work in this
direction is in progress. One can also extend our
approach to the case of folding under quenched force for computing
folding barriers.

We thank M. Kouza for helpful discussions.
MSL highly appreciates the correspondence with M. Rief on AFM experiments.
AMG and AIV are grateful to Kasa im J. Mianowskiego,
Polski Koncern Naftowy ORLEN and Fundacja Zygmunta Zalieskiego for sponsoring
their visits to Warsaw. MSL was supported by
the Ministry of Science and Informatics in Poland
(grant No 202-204-234).

\newpage

\begin{center}
\begin{tabular}{c*{7}{c}}
 & & \; \; $x_{u1}$(\AA)\; \; & \; \; $x_{u2}$(\AA) \; \;&  $\Delta G^{\ddagger}_1/k_BT \;$ & $\Delta G^{\ddagger}_2/k_BT \; $\\
\hline
I27 &\; \; Theory& 3.2 $\pm$ 0.1  &  3.0 $\pm$ 0.1  &  16.9  &  17.0 \\
&\; \; Exp. \cite{Williams_Nature03} &2.2  & 3.0 &20.9&24.2\\
\hline
DDFLN4 &\; \; Theory& 6.1 $\pm$ 0.2  &  5.1 $\pm$ 0.2  &  25.8  &  18.7\\
&\; \; Exp. \cite{Schwaiger_EMBO05,Schlierf_BJ06} &4.0 $\pm 0.4$ & 5.3 $\pm$ 0.4 &17.4&17.2\\
\hline
 \end{tabular}
 \end{center}

\vspace{0.3cm}

Table. Parameters $x_{u1}$, and $x_{u2}$ were obtained in the
Bell approximation. Theoretical values of the unfolding
barriers were extracted from the microscopic theory of Dudko {\em et al}
\cite{Dudko_PRL06}
(Eq. (\ref{Dudko_eq}))
with $\nu = 1/2$, while their experimental estimates were obtained
using Eq. (\ref{UnfBarrier_eq}) and $\tau _A = 0.1 \mu$s.

\newpage

\begin{center}
{\Large \bf Figure Captions} \vskip 5 mm
\end{center}

\noindent {\bf FIGURE 1.}
Schematic plot of
the free energy landscape
for a three-state protein as a function of the end-to-end distance.
$x_{u1}$ and $x_{u2}$ refer to the distance between the
 NS and the first transition
state (TS1)
and the distance between the intermediate state (IS) and the second
transition state (TS2). The unfolding barrier
$\Delta G^{\ddagger}_1 = G_{TS1} - G_{NS}$ and
$\Delta G^{\ddagger}_2 = G_{TS2} - G_{IS}$. A midpoint between TS1 and TS2
$\Delta R^*$ can be determined from simulations.

\vskip 5 mm

\noindent {\bf FIGURE 2.}
(a)
The force-extension curve  at the loading rate
$v=3.6\times 10^7$ nm/s for I27. The results are averaged over 100 trajectories.
The arrows indicate the position of  $\Delta R_{max1} \approx 8$ \AA $\;$ and
$\Delta R_{max2}=78$ \AA.
The vertical solid line refers to $\Delta R^* \approx 43$ \AA.
The inset shows the native state structure of I27
(PDB ID: 1TIT), which contains
seven $\beta$-strands labeled as A to G.
(b) The time dependence of the end-to-end extension for 10 representative
trajectories. Dashed lines refer to the first and the  second plateau, which
are located
at $\Delta R_{p1} \approx 6$ \AA $\;$ and $\Delta R_{p2} \approx 78$ \AA,
respectively. The solid straight line corresponds to
$\Delta R^* = 42$ \AA. $T=285$ K and $f=75$ pN.

\vskip 5 mm

\noindent {\bf FIGURE 3.}
(a) The same as in Fig. \ref{f_ext_v5_snap_titin_fig}a but
for DDFLN4.
The results are averaged over 100 trajectories.
The arrows indicate the position of  $\Delta R_{max1}=14$ \AA $\;$ and
$\Delta R_{max2}=92$ \AA.
The dashed line refers to $\Delta R^* \approx 53$ \AA.
The inset shows the native state structure of DDFLN4
taken from PDB
(PDB ID: 1KSR).
There are seven $\beta$-strands: A (6-9), B (22-28),
C (43-48), D (57-59), E (64-69), F (75-83), and
G (94-97).
In the native state, there are 15, 39, 23, 10, 27, 49, and 20 native contacts
formed by strands A, B, C, D, E, F and G with
the rest of the protein, respectively.
The end-to-end distance in the native state $R_{NS}=40.2$ \AA.
(b) The same as in Fig. \ref{f_ext_v5_snap_titin_fig}b but
for DDFLN4. 
Dashed lines refer to the first and the  second plateau, which
are located
at $\Delta R_{p1} \approx 7$ \AA and $\Delta R_{p1} \approx 85$ \AA,
respectively. The solid straight line corresponds to
$\Delta R^* = 46$ \AA. $T=285$ K and $f=75$ pN.

\vskip 5 mm

\noindent {\bf FIGURE 4.}
(a) The dependence of fractions of native contacts on the end-to-end extension
for DDFLN4. The results were obtained from the
 constant loading rate pulling
simulations with $v=3.6\times 10^7$ nm/s as in Fig. \ref{ddfln4_force_ext_ns_fig}a.
Arrows refer to positions of the peaks in the force-extension curve in Fig.
\ref{ddfln4_force_ext_ns_fig}a.
(b) A typical snapshot at
$\Delta R \approx 14$ \AA. A single contact between
strands A and F is not broken (solid line), while all 11 native contacts
between strands F and G are already broken (dashed lines).
Note that
for the cutoff $d_c=6.5 \AA \;$, there is only one
contact between A and F in the native state.
(c) Shown are probabilities of unfolding pathways
$P_{ufpw}$
for seven $\beta$-strands.
The values of $P_{ufpw}$ are written on top of the histograms.
Results were averaged over 100 trajectories.
\vskip 5 mm

\noindent {\bf FIGURE 5.}
(a) The force dependencies of unfolding times
for $\tau_u$ (circle), $\tau_{u1}$
(squares) and $\tau_{u2}$ (diamonds) for I27 at $T=285$ K.
The straight black lines refer to linear fits
in the Bell approximation for $\tau_{u1}$ and $\tau_{u2}$.
The red curves correspond to the fitting
by Eq. (\ref{Dudko_eq}) with $\nu =1/2$.
The unfolding barriers, followed from this non-linear fit, are listed in
Table. (b) The same as in (a) but for DDFLN4.
The fit curves go up at 
high forces, where the Eq. (\ref{Dudko_eq}) is no longer
 valid \cite{Dudko_PRL06}.

\clearpage

\clearpage

\clearpage

\clearpage

\clearpage


\end{document}